\documentclass[12pt,preprint]{aastex}
\usepackage{lscape}
\usepackage{emulateapj5}           

\begin{document}

\title{High-Mass Proto-Stellar Candidates - II: Density structure from dust continuum and CS emission}
\author{H. Beuther, P. Schilke, K.M. Menten}
\email{beuther@mpifr-bonn.mpg.de, schilke@mpifr-bonn.mpg.de, kmenten@mpifr-bonn.mpg.de}
\affil{Max-Planck-Institut f\"ur Radioastronomie, Auf dem H\"ugel 69, 53121, Germany}
\author{F. Motte}
\email{motte@submm.caltech.edu}
\affil{California Institute of Technology, MS 320-47, Pasadena, CA 91125, USA}
\author{T.K.Sridharan}
\email{tksridha@cfa.harvard.edu}
\affil{Harvard-Smithsonian Center for Astrophysics, 60 Garden Street, MS 78, Cambridge, MA 02138, USA.}
\author{F. Wyrowski}
\email{wyrowski@astro.umd.edu}
\affil{Department of Astronomy, University of Maryland, College Park, USA}
\slugcomment{Version from \today}

\begin{abstract}

We present a detailed 1.2~mm continuum and CS spectral line study of a
large sample of 69 massive star forming regions in very early 
stages of evolution, most of them prior to building up an ultracompact
H{\sc ii} region. The continuum data show a zoo of different
morphologies and give detailed information on the spatial
distributions, the masses, column densities and average densities of
the whole sample.

Fitting the radial intensity profiles shows that three parameters are
needed to describe the spatial distribution of the sources: constant
emission from the center out to a few arcsec radius followed by a
first power law intensity distribution which steepens further outside
into a second power law distribution. The inner flat region is
possibly caused by fragmentation of the large scale cores into smaller
sub-sources, whereas the steeper outer power law distributions
indicate finite sizes of the cores.

Separating the sources into sub-samples suggests that in the earliest
stages prior to the onset of massive star formation the intensity
radial distributions are rather flat resembling the structure of
intensity peaks in more quiescent molecular clouds. Then in the
subsequent collapse and accretion phase the intensity distributions
become centrally peaked with steep power law indices. In this
evolutionary stage the sources show also the broadest C$^{34}$S
linewidth. During the following phase, when ultracompact H{\sc ii}
regions evolve, the intensity power law radial distributions flatten
out again. This is probably caused by the ignited massive stars in the
center which disrupt the surrounding cores.

The mean inner power law intensity index $m_i$ ($I\sim r^{-m_i}$) is
$1.2$ corresponding to density indices $p$ ($n\sim r^{-p}$) of
$1.6$. In total the density distribution of our massive star
formations sites seem to be not too different from their low-mass
counterparts, but we show that setting tight constrains on the density
indices is very difficult and subject to many possible errors.

The local densities we derive from CS calculations are higher (up to
one order of magnitude) than the mean densities we find via the
mm-continuum. Such inhomogeneous density distribution reflects most
likely the ubiquitous phenomenon of clumping and fragmentation in
molecular clouds. Linewidth-mass relations show a departure
from virial equilibrium in the stages of strongly collapsing cores.

\end{abstract}


\section{Introduction}
\label{intro}

In recent years much more attention is being paid to high-mass star
formation. Massive stars release huge amounts of energy in the
interstellar medium, from massive bipolar outflows at their birth to
strong UV radiation throughout their lifetime and terminating
supernovae at their very end, and are thus shaping the interstellar
medium in galaxies. In the past, limited spatial resolution and low
sensitivity made investigations of those regions difficult to
impossible. However, technological advances in the last decade, e.g.,
the advent of mm-interferometers and bolometer-arrays in the mm- and
sub-mm range made it feasible to study even the youngest stages of
high-mass star formation --~prior to ultracompact H{\sc ii} regions~--
in much more detail than it was possible before
(e.g. \citealt{cesaroni 1994,molinari 1998,sridha}). The earliest
stages of High-Mass Proto-stellar Objects (HMPOs) are characterized
observationally by high luminosities, strong dust emission and very
weak or no detectable free-free emission at cm-wavelengths.

Global properties of the sample under study here were presented by
\citet{sridha}. We are now focusing on a detailed analysis of the 
density, mass and column density structure of the dust cores of this
sample of HMPOs. Mapping at 1.2~mm with the IRAM 30~m telescope
reveals the distribution of all dust --~warm and cold~-- with high
enough spatial resolution ($11''$ correspond to 0.1-0.25~pc at typical
distances between 2 \& 5~kpc) to resolve the cores into substructures.
The dust emission at this wavelength is optically thin, and the data
allow determination of the gas and dust column densities, visual
extinction, masses and average densities. These parameters are
important ingredients in the characterization of massive star
formation sites.

Since the sources are resolved the dust maps offer the opportunity of
a structure analysis of the HMPOs, and we present and discuss the
radial intensity distributions of our whole sample. Comparisons of
radial intensity and density profiles with observations of low-mass
star formation sites and with predictions of different star formation
theories (e.g. isothermal or logatropic equations of state,
\citet{shu 1987,mclaughlin 1996}) are essential to understand similarities and
differences between low-mass and high-mass star formation and collapse
processes.

\citep{motte 2001} studied isolated and clustered proto-stellar envelopes, 
and they found that isolated proto-stellar envelopes exhibit density
profiles like $n\sim r^{-2}$ as predicted by the standard model of
star formation by \citet{shu 1987}. In contrast to that, proto-stellar
envelopes in clusters are found to be induced in compact condensations
resembling more finite-sized Bonnor-Ebert spheres than singular
isothermal pheres, which suggests that dynamical proto-stellar models
are more appropriate \citep{motte 2001,whitworth 1985}. Recent
mid-infrared absorption studies of low-mass pre-stellar cores revealed
a flattening of the very inner parts of the radial profiles as well as
a steepening of them further outside \citep{bacmann 2000}. The
high-mass regime is less well observed, but recent studies of
ultracompact H{\sc ii} regions and/or hot cores indicate that their
overall density profiles are somewhat shallower like $n\sim
r^{\alpha}$ with $-1.0<\alpha <-1.5$ than their low-mass counterparts,
as predicted by e.g. logatropic equations of state \citep{vdtak
2000,hatchell 2000}. A physical interpretation of this is that in
low-mass objects thermal support is the main resisting force against
gravitational collapse, while for massive objects non-thermal support
is necessary
\citep{myers 1992,mclaughlin 1996}. It also has to be mentioned that 
earlier studies of ultracompact H{\sc ii} regions found density
distribution of the molecular gas more like $r^{-2}$ \citep{garay
1990,heaton 1993}. Those discrepancies are not finally solved.

We also observed a large fraction of the sample with the 30-m
telescope in various lines of CS and and its isotopomer
C$^{34}$S. Using a LVG code we derive local gas densities and CS
column densities and discuss deviations from the average densities
derived from dust emission. Additionally, the line width information
is used as an evolutionary indicator in correlation with the radial
indices. Further correlations between different parameters based on
dust and CS emission are discussed.

\section{Observations}

\subsection{Millimeter continuum}
\label{obsmm}

The MPIfR 37 element bolometer array MAMBO \citep{kreysa 1998} at the
IRAM 30-m telescope was used to image the 1.2~mm dust continuum
emission of the whole sample. This survey was done in three sessions
between February 1998 and February 2000. The passband of these
bolometers has an equivalent width of approximately 80~GHz and is
centered at $\nu_{\rm{eff}} \sim 240~$GHz.

The images were observed in the dual-beam on-the-fly mapping mode. The
telescope is scanned continuously in azimuth along each row while the
secondary is wobbling in azimuth with a wobbler throw of 50$''$. We
used a scanning velocity of $6''$/sec with a sampling frequency of
2~Hz. Each Scan was separated by $5''$ in elevation. Typical map sizes
with the 37 element array are $300''\times 240''$. Due to being strong
sources, most of the maps were obtained as back-up projects during
rather poor weather conditions (average zenith opacity $\sim
0.3$). The average rms in the final maps is $10-15$~mJy/beam. The
overall calibration uncertainty is estimated to approximately $20\%$.

Data reduction was done mainly with the MOPSI software package
\citep{zylka 1998} and for a few sources with NIC \citep{broguiere 1995}.
Baselines were fitted in time and space. Correlated noise (skynoise)
was subtracted in a radius of $23''$, the flat field for correlated
noise was calculated within the sky noise routine by Zylka \& Haslam
(in prep.).

\subsection{CS observations}

We observed the $J=2-1, 3-2, 5-4$ lines of CS and C$^{34}$S in three
nights between the 31. October and the 3. November 2000 with the IRAM
30-m telescope in a large fraction of sources from our sample remotely
from the MPIfR in Bonn/Germany. The center positions were the
mm-continuum peak positions, and we observed 9 point maps with a $6''$
spacing to map out the beam of the 2--1 line with the 5--4 and 3--2
lines. Calibration errors are estimated to be around
$30\%$. Additional observing parameters are presented in Table
\ref{obs}.

\section {Results}

\subsection{Dust morphologies}

Massive dust cores were detected in all sources of our sample. Figure
\ref{figures} shows all images (absolute {\it IRAS} coordinates are given in
\citealt{sridha}), while Table \ref{mm1} presents the observational
parameters, i.e. the number of identified cores per source (the
numbers in the Table correspond to the labeled numbers in Figure
\ref{figures}), derived peak and integrated fluxes, position
offsets, sizes of the major and minor axis of the sub-sources and the
positional angle (PA), the latter were derived by 2d Gaussian fits to
all sources. The sub-sources were identified by eye because other more
systematic identification criteria (such as 10\% contours or
10$\sigma$ levels) failed in identifying all sources. Integrated
fluxes were determined by sensibly chosen polygons around the
identified sources (and are estimated to be accurate at a $20\%$
level). In the case of isolated sources the polygons correspond to the
$5\%$ levels of the peak emission, whereas for clustered sources, we
had to separate the sub-cores by eye.

We found 154 separate sub-sources in the 69 images, i.e. an
average of 2.2 sub-sources per field. This number is higher than found
in other samples of massive star formation sites, e.g. 1.1
\citep{vdtak 2000}, 1.8 \citep{hatchell 2000} or 1.3 \citep{hunter 2000}. 
This is mainly due to the larger observed fields in our survey
($\approx 16$~arcmin$^2$ in contrast to $4-9$~arcmin$^2$ in the other
samples), the number of sources per arcmin$^2$ does not vary strongly
throughout the different samples. Most likely the majority of the
single-dish cores split up into a number of sub-cores again at even
higher resolution because of the clustered mode of massive star
formation.

Morphologically the sources resemble a zoo of different features:
while many are centrally peaked, massive cores (e.g. 05553+1631,
18264-1162) with maybe smaller sub-sources nearby (e.g.18182-1433,
19266+1745) others show multiple associated, massive cores
(e.g. 18151-1208, 18310-0825), elongated structures (e.g. 18102-1800,
20081+2720), larger filaments (e.g. 18223-1243, 19175+1357) and a few
with less peaked features (e.g. 18540+0220, 19282+1814). We refrain
from a systematic morphological classification, because many features
are subject to resolution and projection effects, and a formal
classification would be misleading. Nevertheless, the diversity of
sources offers the chance to study many different aspects of massive
star formation at its earliest evolutionary stages and its evolution
in time: are the less peaked sources younger than the peaked ones, is
the density structure different from the low mass case and if so, what
does that imply, how do different sources interact with each other?

As can be seen from Table \ref{mm1}, the large majority of sources is
resolved by the $11''$ beam. However, higher resolution
interferometric observations of a few sources (Beuther et al., in
prep.) reveal that the cores contain sub-structures which are not
resolved by the single-dish observations. Combining both observations
indicate that central, small condensations exist, which are embedded
in larger cores observed at lower angular resolutions.

Other features of the dust cores --~molecular line emission, maser
emission, weak cm emission~-- are discussed in \S \ref{subgroups} and
\citet{sridha}.

\subsection{Column densities, visual extinction, masses and average densities}

The 1.2~mm continuum is mainly due to emission from optically thin
dust \citep{hildebrand 1983}. Following the approach outlined in
\citet{hildebrand 1983} we calculate the beam-averaged gas column
density and the mass of the cores to

\begin{eqnarray*}
M_{\rm{gas}} & = & \frac{1.3\times10^{-3}}{J_\nu(T_{\rm{dust}})} 
\frac{a}{0.1\mu \rm{m}} \frac{\rho}{3 \rm{g cm}^{-3}} \frac{R}{100} 
\frac{F_{\nu}}{\rm{Jy}}\\
             &   & \times \left(\frac{D}{\rm{kpc}}\right)^2 
      \left(\frac{\nu}{2.4 \rm{THz}}\right)^{-3-\beta}\ [\rm{M}_{\odot}]\\
N_{\rm{gas}} & = & \frac{7.8\times10^{10}}{J_\nu(T_{\rm{dust}})\Omega} 
\frac{a}{0.1\mu \rm{m}} \frac{\rho}{3 \rm{g cm}^{-3}} \frac{R}{100} 
\frac{F_{\nu}}{\rm{Jy}}\\
             &   & \times \left(\frac{\nu}{2.4 \rm{THz}}\right)^{-3-\beta} 
    \ [\rm{cm}^{-2}]
\end{eqnarray*}

where $J_\nu(T_{\rm{dust}})=[\exp(h\nu/kT_{\rm{dust}})-1]^{-1}$ and
$\Omega, a, \rho$, $R$ and $\beta$ are the beam solid angle, grain
size, grain mass density, gas-to-dust ratio and grain emissivity index
for which we used the values 0.1$\mu m$, 3g~cm$^{-3}$, 100 and 2,
respectively \citep{hunter 1997,hunter 2000}. The dust temperatures
$T_{\rm{dust}}$ (ranging between 30~K and 60~K) are obtained by
greybody fits to the {\it IRAS-} and mm-data
\citep{sridha}. Systematic errors in these quantities are prevalent
for the mass determination, e.g. the grain emissivity index $\beta$ is
reported to vary at least between 1.75 and 2.5 between the Orion
photon-dominated region and the Orion ridge
\citep{lis 1998}, but work on massive star formation regions
\citep{hunter 1997,molinari 2000} supports the use of a value of 2
on the average for such objects. Higher dust opacity indices as proposed by  
\citet{ossenkopf 1994} would result in masses and column densities about 
a factor 4 lower, whereas lower temperatures --~as proposed by NH$_3$
observations \citep{sridha}~-- would increase the derived parameters
more than a factor of two. The approach outlined above is used by us
because recent observations of ultracompact H{\sc ii} regions
\citep{hunter 1997, hunter 2000} were analyzed similarly, and 
a consistent data analysis is essential when comparing different
observations. Taking into account all the possible errors and
uncertainties we estimate the mass and column density to be correct
within a factor of 5. Additionally, we calculated the visual
extinction A$_v=N_{\rm{gas}}/0.94\times 10^{21}$
\citep{frerking 1982} and the average density n$_{\rm{gas}}$ by
dividing the column density by the average of the major and minor
source size at the given distances
\citep{sridha}. All the derived quantities are presented in Table
\ref{mm2}. A comparison of the masses and luminosities of this sample 
with ultracompact H{\sc ii} regions is presented in \citet{sridha}.

Beam averaged column densities of the main clumps range around a few
times 10$^{23}$~cm$^{-2}$, which correspond to a few 100 magnitudes of
visual extinction. There are three exceptions (18089$-$1732,
18182$-$1433 and 18264$-$1152) with extremely high column densities as
large as 10$^{24}$~cm$^{-2}$ corresponding to a visual extinction of
$10^4$. These sources are prominent from our other observations as
well \citep{sridha}: they show no cm-emission down to 1~mJy, have
water and methanol maser emission and are strong in all molecular
lines we observed so far, outflow tracers as well as dense gas
tracers. Many of the secondary clumps have column densities below
10$^{23}$~cm$^{-2}$, which might correspond to lower mass star
formation sites associated and interacting with the main massive
cores.

While the column density and correspondingly the visual extinction are
distance independent quantities the derived densities and masses
suffer from ambiguities basing on the kinematic distance determination
\citep{sridha}. Therefore Table \ref{mm2} gives always a far and a
near value for masses and densities (if just a far value is listed, we
could resolve the distance ambiguity). The masses of the main cores
are very high ranging from hundreds to a few thousand M$_{\odot}$, and
for the far distance masses of $10^4$~ M$_{\odot}$ are reached. The
most massive cores correspond to the sources with the highest column
densities. The masses of the adjacent sub-cores were calculated using
the same T$_{\rm{dust}}$ as for the main cores, but it is possible
that those sub-cores are colder, which makes us underestimating the
masses and column densities. Future independent temperature
determinations of each sub-core are of great interest.

Additionally, we derived the average densities of the cores, and again
we see differences between the main cores and the secondary
cores. Average densities of the main components are mostly a few times
10$^{5}$~cm$^{-3}$, while the less massive secondary clumps are about
one order of magnitude less dense.

\subsection{Radial Profiles}

\subsubsection{Observational limitations of the intensity distributions} 
\label{filtering}

The maps are a two-dimensional intensity distribution, which is a
convolution of the intrinsic intensity distribution with the telescope
beam and which is also affected by the mapping technique. The
intrinsic intensity distribution is a function of the density and
temperature fields of the source, quantities which we ultimately want
to derive. To interpret our results we have to consider carefully how
all these factors affect the observations.

The intensity distribution of a source is convolved with the beam of
the telescope, which limits the smallest observable spatial scale down
to the HPBW. But theoretical work \citep{adams 1991} as well as
simulations \citep{motte 2001} show that the convolution of the
intensity distribution with the beam does not affect scales larger
than the beam significantly.

However, effects of the dual-beam mapping technique \citep{ekh} and
the size of the array have to be taken into account when analyzing
larger scale source structures. \citet{motte 2001} showed that part of
the emission is filtered out due to the dual-beam mapping technique if
there is a gradient of emission at the edge of the map. For most of
our maps the emission decreases below the noise already far away from
the edge of the map, which causes an ambiguity in the sense that we do
not know whether the sources extend outside the maps below our
sensitivity limit or whether the sources are fully contained in our
mapped area and maybe just embedded in a constant low level plateau of
the surrounding cloud. In the latter case of a constant plateau no
spatial filtering effects the data. But as spatial filtering due to
low level emission of the sources might steepen the radial profiles
slightly, we simulated observations (similarly to those simulations
outlined in
\citealt{motte 2001}) for our map sizes and found that the possible 
steepening due to spatial filtering is largely compensated by
flattening effects caused by the error beam lobe of the 30m
telescope. In any case, the power law index of the radial intensity
profile measured over radii smaller than $50''$ only needs to be
corrected by values $<0.2$. Regions outside this radii still give
information about morphology, number of sub-sources etc., but the
absolute profiles there have to be taken more cautiously.

Hence in our analysis we concentrate on the range of radii between
$11''$ and $50''$ from the center since that should reflect the
intrinsic radial profiles reasonably well. Structure on scales smaller
than the HPBW of $11''$ cannot be resolved, but the data constrain the
integrated flux.

\subsubsection{Radial intensity profiles and fitting results}
\label{fitting}

As already mentioned in the Introduction, different star formation
theories predict different density distributions of the core
envelopes: e.g. singular isothermal spheroids result in proto-stellar
density power law distributions $n \sim r^{-2}$ \citep{shu
1987,whitworth 1985}, whereas logatropic equations of state predict
flatter distributions $n \sim r^{-\alpha}$, with $1.0$
\citep{mclaughlin 1996}. Therefore, to probe theories we try to fit
our data by power law distributions.

Although the observed maps show substructure and multiple sources are
a common phenomenon, most of the clumps themselves are spherical to
first order. But non-circular deviations most likely correspond to
substructures in the core center than to changes in power law
distributions, because comparing power law distribution s for cuts at
various position angles generally produces similar power law
indices. Therefore we produce the radial profiles by averaging the
observed intensities in circular annuli of $4''$ (slightly less than
Nyquist sampling) around the peak position. If adjacent and secondary
sources distort the circular averaging significantly those ranges are
not considered for fitting profiles. Fitting the data between $12''$
and $52''$ by one power law gives poor results. Introducing a break at
$32''$ (half of the fittable range) and fitting the data by two power
laws reveals that in most cases the two power law fits are better,
i.e. they have a far lower $\chi ^2/\nu$ (sum of squared deviations
between fit and data divided by the degrees of freedom) than the
single power law fits. 

This basically says that power laws are not a very good description of
the structure, but for simplicity's sake and for comparison with
theoretical models, we still fit the data in these terms. We are aware
that the break at $32''$ is arbitrary, because of the different
distances and linear sizes of the sources, but we choose this approach
for practical reasons. Introducing another varying break parameter
would increase the degrees of freedom further, which is not desirable
with altogether only 10 useable spatial positions. The exact location
of the outer break is not crucial to the result that a single power
law fit is inadequate. We also stress that while the exact values of
the power law indeces do depend on the particular location of the
outer break, our qualitative result, the steepening outwards, is not
affected by the exact location. Combining all intrinsic errors (noise,
steepening of the profiles by spatial filtering, flattening of the
profiles due to the error beam, calibration, fitting error etc.) we
estimate the $\sigma_m$ of the radial profiles $I\sim r^{-m}$ to
$\Delta m \sim 0.3$ (see also \citealt{motte 2001}). The inner and
outer profiles (all inside the radii of $\sim 60''$) are affected very
similarly by the observational problems outlined in \S
\ref{filtering}.

Our fitting procedure uses 3 parameters: the outer region from $32''$
to $52''$ is fitted by one power law $I\propto r^{-m_o}$ followed by a
power law $I\propto r^{-m_i}$ out to $32''$. Since the flux is finite,
this power law has to stop somewhere, and we model the central region
with constant intensity inside a breakpoint $b$.  Those models were
convolved with the beam of $11''$ and $b$ and $m_i$ were fitted
simultaneously to the data, whereas $m_o$ was fitted directly to the
data, because at those scales the effect of the beam is negligible.

The results are presented in Table \ref{mm2} and a few example fits
are shown in Fig \ref{fits}. Figure \ref{histo_profiles}(a) presents
the distribution of inner radial indices $m_i$ with a mean of 1.2 and
a broad peak around 1.4. Fewer outer fits were possible due to
secondary sources or too weak emission in that regions. An even better
determination of the outer power law indices requires observing maps
of a far larger extent, but the steepening in the outer region is a
significant feature in a large fraction of sources. Approximately
$65\%$ of the sources, where $m_o$ was determinable, show a steepening
$m_o-m_i \geq 0.4$ ($\sigma_m \sim 0.3$) with an average $m_o-m_i \sim
0.6$ (see Fig. \ref{histo_profiles}(c)). While the $1\sigma$ error of
an individual determination is 0.3, the large number of sources make
the qualitative result of a significant steepening outwards
reliable. We also want to point out that in only one case (18553+0414)
we find a flattening towards the outer core.  In the case of low-mass
sources a steepening in the outer region has recently been shown by
\citet{bacmann 2000} and \citet{alves 2001} with the totally different 
techniques of mid-infrared and near-infrared absorption studies,
respectively, which both do not have spatial filtering effects.

\subsection{CS and C$^{34}$S observations}

We observed CS and C$^{34}$S in 89 clumps (54 sources) and detected
the CS(2--1) and (3--2) line in all of them. This is not a surprise
because one of the original sample criteria was a detection in the
galaxy-wide IRAS-CS survey by \citet{bronfman 1996}. In a few sources
we found previously unknown secondary velocity components (see bottom part
of Table \ref{table_cs}). The CS(5--4) line and the corresponding
C$^{34}$S lines were found in large sub-samples (Table
\ref{table_cs}). A few sources show self-absorption effects 
($18089-1732$, $18310-0825$, $18337-0743$, $18460-0307$, $18488+0000$
\& $19217+1651$) in the main isotopomer CS, but all other lines can be
fitted well by Gaussian profiles. Table \ref{table_cs} presents the
systemic velocity of each set of lines, the peak temperatures
$T_{\rm{mb}}$, the integrated intensities $\int T_{\rm{mb}}dv$ and the
line widths $\Delta v$ of all detections. \\

We use a LVG-code to determine local gas densities and CS column
densities within a beam of $\sim 22''$. For this purpose we use only
sources for which 9 point maps were available to smooth the CS(5--4)
and CS(3--2) lines spatially to the resolution of the (2--1)
line. Additionally, we use for the LVG-calculations only sources with
at least 3 different detected lines, because only line ratios can be
used to determine physical parameters since the absolute line
strengths are very unreliable due to the unknown beam filling
factor. 61 sources fulfill these criteria for reasonable density and
column density estimates. \citet{schilke 2001} note that for diatomic
molecules line ratios cannot be used to determine the density or
temperature independently but that the pressure $p=n\cdot T$ is
constrained very well. We therefore calculate the LVG-models for 20,
40 and 80~K with gas densities of $10^4-10^6$~cm$^{-3}$ and CS column
densities of $10^{12}-10^{16}$~cm$^{-2}$. We then performed $\chi^2$
fits of our data to the LVG-model and calculated the incomplete gamma
function as a goodness of fit estimator \citep{press 1996}. The
$3\sigma$ levels of the incomplete gamma function were taken to derive
the range of parameters our observations can be fitted with. Table
\ref{mm2} presents the derived thermal pressure $p$ and the CS column
density $N$(CS) for temperatures between 20 and 80~K. For cases where
only single values are given the $3\sigma$ level of the incomplete
gamma function revealed parameter ranges with errors of less than
$10^{0.25}$. \citet{sridha} showed that temperatures derived from {\it
IRAS} infrared observations are in the range of $\sim 40$~K while the
rotation temperatures of NH$_3$ trace colder material around
20~K. Additional CH$_3$CN observations in a sub-sample indicate higher
temperatures around 100~K.  We estimate the gas density by assuming an
average temperature of 40~K for the sample. The last two columns of
Table \ref{mm2} give the gas density range and CS column density range
of possible parameters at 40~K.

For further statistical comparison we derive discrete parameters for
the column and volume densities at 40~K: we take the mean value if
ranges are given, the upper limits for CS column densities if
C$^{34}$S is a non-detection, and the lower limits in the density
distribution if no upper limits for the gas density can be
derived. To derive meaningful column densities those values have to be
multiplied with the linewidth (C$^{34}$S 2--1 if observed, else CS
2--1), because the LVG code gives column densities per km/s.

Figure \ref{histo_cs}(a) and (b) presents the gas density and CS
column density distribution for the 61 sources we used in the
calculations. The average local gas density is $1.0\times
10^6$~cm$^{-3}$ and the average CS column density is $1.3\times
10^{15}$~cm$^{-2}$. Additionally, we find that the CS abundance is
independent of the core mass with an average value around $8\times
10^{-9}$.

\section{Discussion}

While the average gas densities we derive from the dust continuum
emission are around $10^5$~cm$^{-3}$ the local gas densities implied
by the CS observations are larger (up to one order of
magnitude). Inhomogeneous density distributions are not unexpected
because molecular clouds are clumped in all stages of molecular cloud
evolution, from quiescent clouds in a pre-collapse phase
(e.g. \citealt{beuther 2000}) to dense massive cores we are observing
here (see also \citealt{stutzki 1989}). Because CS is sensitive to the
denser gas, the difference in both density determinations could also
be explained simply by a density gradient of the core. However, as
massive star formation is generally considered to proceed only in a
clustered mode \citep{stahler 2000} we favor the clumping as the most
likely explanation (see also \S \ref{breakinner}). Comparing our sample
with hot cores and ultracompact H{\sc ii} regions UCH{\sc ii}s
(e.g. \citealt{kurtz 2000}) shows that HMPOs, UCH{\sc ii}s and hot
cores resemble each other in masses and column densities.

\subsection{Implications from the radial profiles}

It has to be stressed that in many sources the intensity profiles are
not well fitted by single power laws (\S \ref{fitting}). Our approach
using 3 parameters describes the intensity distribution significantly
better: a steeper outer power law, a flatter inner power law and an
inner breakpoint.

It has to be noted that although most descriptions of density profiles
are expressed in terms of single power laws in low mass star formation
research (\S \ref{intro}), \citet{ward 1999} find centrally flattened
density profiles for a significant number of pre-stellar cores being
in a state of evolution prior to protostars, whereas \citet{alves
2001} see a steeper outer region in one Bok globule modeled by
Bonnor-Ebert spheres. Additionally, \citet{motte 2001} found finite
sizes for proto-stellar envelopes within clusters. A recent
mid-infrared absorption study by \citet{bacmann 2000} revealed that
inner flattening and outer steepening of pre-stellar low-mass cores
occurs regularly. \citet{henriksen 1997} model pre-stellar cores by a
centrally flattened hydrostatic core, an inner flatter power law
distribution which is followed by a steeper outer power law
distribution (essentially Bonnor-Ebert spheres), very similar to the
results we obtain from our fits. In spite of the overall similarity of
their model assumptions and our fitting results this is probably a
chance coincidence since our sources most likely are cluster, whereas
\citet{henriksen 1997} model single sources within clusters. But 
we like to stress that the intensity profiles \citet{bacmann 2000}
derive for their low-mass pre-stellar cores are qualitatively similar
to our results in the high-mass regime.

In the massive star formation regime \citet{bonnell 1998} model the
accreting cores with shallow inner density distributions embedded in
steeper outer regions. The outer steepening of the radial profiles is
most likely due to the finite size and by that the finite
mass-reservoir of the cores \citep{bacmann 2000}. \citet{heaton 1993}
also found an inner region of constant molecular hydrogen density
associated with the ultracompact H{\sc ii} region G34.3+0.2 (see also
\S \ref{breakinner}). While in the low-mass regime there do exist a
number of theoretical models predicting different pre-stellar
intensity distributions than just one power law (e.g. \citet{bonnor
1956,basu 1994,henriksen 1997}), theoretical high-mass research in
that direction --~especially the evolution of massive cores giving
birth to whole cluster~-- has been rare so far and should be
intensified in the future.

\subsubsection{Different sub-groups and evolutionary effects}
\label{subgroups}

The distribution of inner radial profiles $m_i$
(Fig. \ref{histo_profiles}(a)) is broad, and we try to distinguish
between the different parts of that distribution. Comparing the radial
indices with other parameters such as luminosity, molecular
observations \citep{sridha}, the mass and the CS-column density, there
are indications that the steeper indices correlate with the more
luminous and more massive sources with high column densities, which
show hot core signatures in molecular line observations and/or maser
emission \citep{sridha}. The flatter sources might be divided into two
sub-groups: one has adjacent resolvable cm-sources \citep{sridha}
indicating a more evolved state of evolution \citep{sridha}, where the
evolving star has already disrupted the core (e.g. 22551$+$6221). The
other sub-group with flat distributions is on the average less
prominent in other tracers, they are less massive, mostly adjacent to
the main cores, have lower column densities and show no other
prominent star formation signature, particularly no cm-continuum.

To strengthen this hypothesis we select three sub-samples and present
them in Fig. \ref{histo_profiles}(b): the first sub-sample shows
CH$_3$OH and/or CH$_3$CN emission and has no or only hypercompact
unresolved cm-emission features, the second sub-group shows resolvable
($>2''$) ultracompact H{\sc ii} regions, and the third consists of
CH$_3$OH and CH$_3$CN non-detections, which have no or only
hypercompact unresolved cm-emission (the molecular and cm data are
taken from \citet{sridha}). The different groups are indicated in the
last column of Table \ref{mm1}. We refrained from classifying adjacent
sub-cores and those where no good molecular line or cm data were
available.  We are aware of the low-number statistic, thus we doubled
the bin-size to increase the statistical significance. The first
group, consisting of hot core type sources, has on the average steeper
radial indices $m_i$ (mean$\sim 1.5$) compared to the resolved
cm-sources and the molecular line non-detections, which both have a
mean radial index of $\sim 1.1$.

We interpret the different sub-groups in Fig \ref{histo_profiles}(b) in
an evolutionary scenario: at the very early stages before or just at
the beginning of star formation the profiles are rather flat with less
pronounced gravitational centers and no or just weak internal heating
(accretion and collapse begins). During the subsequent collapse and
accretion phase before or at the very beginning of nuclear fusion the
profiles are steepest due to strong accretion onto the central
sources. After the first massive stars have formed in the cluster,
they inject so much energy in the surrounding medium that the cores
disrupt, which is observable by again flattening intensity
distributions. Our sample consists of sources in all these stages and
is therefore perfectly suited for further studies of different
evolutionary stages.

\subsubsection{The inner breakpoint}
\label{breakinner}

The third parameter we use to characterize the radial intensity
profiles is the inner breakpoint of the power laws, from where on we
model the emission on a constant level to the center. Similarly,
\citet{heaton 1993} modeled the density distribution of the molecular 
gas associated with the ultracompact H{\sc ii} region G34.3+0.2 based
on HCO$^+$ and H$^{13}$CO$^+$ observations as approximately constant
over the inner 0.1~pc ($\sim 5.5''$ at 3.8~kpc). For 2 sources
(05358+3543 \& 19217+1651) we have interferometric mm-continuum data
from Plateau de Bure with approximately $3''$ resolution (to be
published in a forthcoming paper, Beuther et al. 2001b, in prep).  The
massive cores are resolved into sources of sizes $\leq 5''$, and the
nearby source 05358+3543 fragments into at least 3 massive sub-sources
(distance $\sim 1.8$~kpc versus $\sim 10$~kpc for 19217+1651). The
correspondence between the size of the sub-sources and the breakpoints
of the power laws indicates that the breaking of the power laws is due
to fragmentation of the cores. Each fragmented sub-core has its own
substructure, but on a $11''$ scale, where we just see the emission
integrated over the sub-cores, this appears as flattening. We
investigated whether this flattening could be caused by increasing
optical depth at small scales by integrating density power laws
(derived from the intensity profiles, see section
\ref{density_profiles}) along each line of sight, and find that the
sources stay optically thin into the sub-arcsecond regime. This makes
opacity effects very unlikely to explain the breakpoints.

Figure \ref{break_mass} plots the corresponding linear distances from
the peak position $b_{\rm{lin}}$ versus the derived core mass for
sources without distance ambiguity. The linear distance from the peak,
where the power law breaks increases with increasing mass (fitting
results in approximately $b_{\rm{lin}}\propto M^{0.6}$). This seems to
be plausible in the fragmentation scenario as well, because the more
massive the cores are the larger and more massive the evolving
clusters are expected to be (see also \citealt{sridha}), which leads
to fragmentation on larger scales. 

We excluded for the fit the three sources at the bottom (open circles) of
Figure \ref{break_mass} ($05553+1631$, $18151-1208$ and $19471+2641$),
because they clearly show a different behavior with a breakpoint
below $0.5''$, which means literally no breakpoint at all in our
description. In the case of $19471+2641$ cm-imaging snapshot-data
\citep{sridha} are very bad, but we think that extended emission is 
the cause of the data problems. This can explain the missing
breakpoint easily because an ultracompact H{\sc ii} region has already
destroyed the core significantly, which is also indicated by the very
filamentary structure of our 1.2~mm image. It is different in the
cases of $05553+1631$ and $18151-1208$, which show both many signs of
very early stages of massive star formation \citep{sridha}. Therefore
we think that those sources are dominated strongly by one massive
central objects, which makes fragmentation undetectable at our scales.

\subsubsection{Density profiles}
\label{density_profiles}

As outlined so far single power laws do not fit the intensity profiles
well, and therefore the derivation of single power law density
profiles seem not to be very reasonable. Nevertheless, it is
interesting to estimate density profiles $n \propto r^{-p}$ at
least for the inner intensity power laws and compare those with
theoretical predictions and observational studies, e.g. \citet{osorio
1999,motte 2001,hatchell 2000,vdtak 2000}. For optically thin dust
emission, assuming the Rayleigh-Jeans approximation, roughly spherical
cores and temperature and density distribution following a power law,
the intensity index $m$ depends to first order on the density index
$p$ and the temperature index $q$ ($T
\propto r^{-q}$) via \citep{motte 2001,adams 1991}:

\begin{eqnarray}
m=-1+p+Q\cdot q+\epsilon _f
\label{m}
\end{eqnarray}

Q is a temperature and frequency dependent correction factor and
equals $\sim 1.2$ at 1.2~mm and 30~K. The original equation \ref{m}
was derived for infinite power law distributions, but our steeper
outer power laws indicate for a fraction of sources finite
sizes. Those finite sizes have to be taken into account by a
deprojection correction $\epsilon _f$ \citep{yun 1991}, but
\citet{motte 2001} showed that $\epsilon _f$ is small, roughly
$\epsilon _f \leq 0.2$ if the outer edge is $\geq 20\%$ of the map
size and even $\epsilon _f \leq 0.1$ if the outer edge is $\geq 30\%$
of the map size. In spite of possible errors this approach gives a
first estimate on the density distribution. To derive the density
index $p$ we therefore need some knowledge about the temperature
distribution. In the isothermal case the temperature does not depend
on the distance to the peak, but in our cases of assumed centrally
heated sources (accretion and/or stellar burning), the temperature
decreases with distance to the source and
\citet{emerson 1988} shows that the power law index $q$ of the
temperature distribution can be approximated reasonably well by
0.4. We have additional 870$~\mu$m maps for 13 of the sources, and
comparison of the 870$~\mu$m and the 1.2~mm fluxes backs up the idea
that the heating stems from the inside, at least for the main
cores. Simulations of massive star forming regions
\citep{osorio 1999,vdtak 2000} revealed that in the inner regions
(approximately $<2000$~AU) steeper indices are expected. But outside
the inner 2000~AU the temperature distribution flattens again and runs
asymptotically into the usual $r^{-0.4}$ distribution \citep{vdtak
2000}. With our resolution being typically between 20000 and 50000 AU
we can safely assume this asymptotic solution with $q=0.4$ and derive
the density power-law index $p=1.6$, while the peak is at $p=1.8$. As
outlined by \citet{motte 2001} we estimate the error in $p$ to $\Delta
p \sim \Delta m+\Delta q + 0.1 \sim 0.5$ with $\Delta m\sim 0.3$ and
$\Delta q \sim 0.1$. We checked the derived density and temperature
distributions by several DUSTY \citep{ivezic 1999} and DIRT
\citep{wolfire 1986} radiative transfer program runs and got similar
results to the straightforward deviation described above. The sources
with very low $m$ might not be centrally heated, yet, and by that a
near isothermal temperature distribution is more likely ($q<0.4$),
thus their density indices are probably slightly larger.

Dividing the sample again in the same sub-groups of strong molecular
emission sources, weak molecular emission sources and extended
cm-emission sources (see section \ref{subgroups}), the strong
molecular sources have a mean density index of 2.0, the weak molecular
sources 1.6 and the cm-sources 1.5.

As already mentioned, the standard theory of low-mass star formation
from singular isothermal spheres \citep{shu 1977} predicts a power law
index for the density distribution of $r^{-2}$ in the static outer
proto-stellar envelope. Theorist of high mass star formation propose a
different, logatropic equation of state \citep{osorio 1999,mclaughlin
1996} to explain e.g. the linewidth-size relation observed in
molecular clouds. Those logatropic equations of state result in a
different density distribution in the outer envelope, namely $n
\propto r^{-1}$. The physical interpretation of the logatropic
equation of state is under debate, but it is argued that in low mass
star formation sites mainly thermal support works against collapse,
while in massive sites non-thermal support is necessary. Recent
observational work in molecular cores associated with well known
ultracompact H{\sc ii} regions (UCH{\sc ii}s) and hot cores argue in
favor of density indices around 1.5 \citep{hatchell 2000,vdtak 2000}.
But there do also exist observations of molecular gas associated with
ultracompact H{\sc ii} regions, which favor density distributions
similar to the low-mass case: e.g. independent studies of G34.3+0.2 in
NH$_3$ and HCO$^+$ found density distributions around $r^{-1.9}$,
respectively \citep{garay 1990,heaton 1993}.

The mean density index $p=1.6 \pm 0.5$ derived for the inner density
profile in our sample does not support strong deviation from the
low-mass proto-stellar profiles studied by \citet{motte 2001}. The
main difference between both studies is that \citet{motte 2001} derive
profiles for isolated low-mass proto-stellar envelopes and low-mass
proto-stellar envelopes within clusters, while our sample presents the
envelopes of massive star forming clusters. The evolutionary effects
outlined in \S
\ref{subgroups} with steepening density index in the strongest
collapse and accretion phase and again flattening profiles, when the
first ignited massive stars disrupt the molecular cores, indicate that
with ongoing star formation turbulence gets more and more
important. The steepest indices during the collapse and accretion
phase resemble very much the structure known from isolated low-mass
sources.

The density indices derived by \citet{hatchell 2000} and
\citet{vdtak 2000} for well known hot cores associated with
ultracompact H{\sc ii} regions favor more strongly the logatropic
equations of state ($1.0<p<1.5$). They seem to correspond well with our
flatter indices observed in more evolved cm-sources. To check that we
applied our fitting procedure to the sources observed by
\citet{hatchell 2000}\footnote{The data were kindly provided by
Jennifer Hatchell.}. The average intensity index $m$ derived by us for
those hot cores from $13.5''$ to $34.5''$ from the peak position is
1.5, resulting in density indices $p$ around 1.9, which puts them in a
similar category as the peaked sources of our sample and shows that
the ultracompact H{\sc ii} regions have not disrupted the molecular
cores significantly. The differences between the density indices
derived in this two different ways, indicates how difficult it is to
really constrain density distributions. The discrepancy might be
explained to first order because \citet{hatchell 2000} describe their
data by single power laws, while we argue that this is not sufficient.
Especially, including the radial positions inside the beam into single
power laws flattens the profiles because the power laws break at
distances of a few arcsec from the center and the integrated emission
continues more or less flat. But it has to be noted that
\citet{hatchell 2000} fitted simultanously the SEDs, which puts additional 
constraints on the radial distributions.

We conclude that the analysis of density distributions is very
dependent on model assumptions, and that comparisons of different data
sets are only meaningful if they are analyzed using the same
methods. Future approaches should try to incorporate different power
laws or maybe Bonnor-Ebert spheres to describe the envelopes of
massive star forming regions more properly, but that is out of the
scope of this paper.

\subsection{Line-width relations}
\label{linewidths}

The FWHM of Gaussian fits to the observed profiles give line
information about the state of turbulence and the region where the gas
is mainly emitting. Figure
\ref{widths}(a) plots the FWHM of C$^{34}$S(2--1) versus NH$_3$(1,1) 
and (2,2) linewidth (filled and open circles for a sub-sample of our
HMPOs without distance ambiguities, see \citealt{sridha}) and a few
molecular cores associated with UCH{\sc ii} regions \citep{churchwell
1990,cesaroni 1991}. The asterisks and triangles represent the NH$_3$
(1,1) and (2,2) lines, respectively. The HMPOs generally show a
broader linewidth in C$^{34}$S than in NH$_3$ with an average ratio of
$\frac{\Delta v(\rm{C^{34}S})}{\Delta v(\rm{NH}_3)}\sim 1.5$. Similar
linewidth differences are found in low mass dense cores
\citep{zhou 1989}. Mapping of both lines by the same authors revealed
that the average spatial extension of CS is larger than that of
NH$_3$, which supports well known linewidth-size relations
\citep{larson 1981}. This is somewhat surprising since NH$_3$ traces 
less dense material than CS. For dark cores the effect seems to be due
to the fact that CS is depleted in the dense cores while NH$_3$ is
not, so the former traces the extended, more turbulent envelope, while
the latter mostly traces the only thermally supported dense core. In
our case, freezing out of CS is unlikely because of our high average
temperatures, also our size scales are different. We propose that in
our sources NH$_3$ traces the more quiescent medium density interclump
gas, while CS emits mostly from the star forming and hence more
turbulent cores.

The comparable database of 7 hot cores associated with UCH{\sc ii}
regions \citep{churchwell 1990,cesaroni 1991} is small but shows that
the average linewidth of these hot cores is significantly higher than
that of HMPOs (see also \citealt{sridha}). In few of the UCH{\sc
ii}/hot core regions, where the linewidths are comparable, the ammonia
abundance seems to be enhanced by grain mantle evaporation in hot
cores, so NH$_3$ might trace those regions as well.

{\sl How does the linewidth correlate with mass of the cores and state
of evolution?} Figure \ref{widths}(b) presents the C$^{34}$S(2--1)
linewidth at FWHM versus the core mass derived from the dust
continuum data. The full circles show the HMPOs without distance
ambiguity, while the asterisks and open squares present data from
well known UCH{\sc ii} region associated hot cores. The latter C$^{34}$S
linewidth are taken from \citet{cesaroni 1991}, and the masses for
the hot cores are calculated from C$^{34}$S data (asterisks,
\citealt{cesaroni 1991}) and from dust continuum or C$^{17}$O
observations \citep{hunter 1997,hatchell 2000,hatchell 1998}.  As full
line we show the relation which is expected in case of virial
equilibrium (see e.g. \citealt{myers 1999}). While three hot cores are
at the upper linewidth-end of the HMPOs, the majority of them lies far
nearer the virial equilibrium track. Conversely, in our sample only the
less massive objects correlate with the expected virial equilibrium
correlation, whereas the more massive objects have significantly
narrower linewidth and cluster in the plot below the hot core
region. From an evolutionary point of view this seems to indicate that
cores stay in virial equilibrium before the cores start collapsing,
then with beginning star formation the whole massive core leaves the track
of virial equilibrium and collapses violently until the first
(massive) stars ignite and stabilize the overall collapsing motion at
least partly by increasing the total luminosity. Accretion should still
be possible at that stage of evolution but the overall collapse of
the core is most likely stopped or at least slowed down. This effect
of departure from the virial equilibrium track seems to be more
prominent the more massive the cores become.

\section{Conclusions}

The presented database of 69 1.2~mm continuum maps of a homogenous
sample of massive star-formation regions at very early stages
of evolution sets many constraints on the physical parameters of
high-mass proto-clusters.

The maps help disentangle many substructures, and it is obvious that
a zoo of different morphologies is present. Calculations of total
masses, column densities and densities show that HMPOs and hot cores
associated with ultracompact H{\sc ii} resemble each other with regard
to those parameters.

An analysis of the radial intensity profiles reveals that single power
laws do not fit the data well, but that three parameters describe the
radial profiles qualitatively better: a steeper outer power law, a
flatter inner power law and an inner breakpoint, where the power law
description stops, and we model the distribution as flat.

Larger scale maps will help to set better constraints on the outer
envelope structure, but we propose that this steepening is due to the
finite size of the cores as proposed in low-masses pre-stellar cores
\citep{bacmann 2000}. Detailed theoretical models of the dust and gas 
envelopes of massive star forming regions are needed in the
future. Comparing the inner breakpoints of the power law distributions
with interferometric observations at higher resolution indicates that
the breakpoints correspond to the separation and/or sizes of smaller
sub-sources. Thus, most likely fragmentation causes the breaking of
the power laws. Each individual sub-core might show other power law
distributions, but this has to be investigated with much higher
resolution (Beuther et al. in prep., Wyrowski et al. in prep).

Regarding the inner power law indices, we can divide the sources into
three sub-samples: sources with strong molecular line emission,
sources without strong molecular line emission or other strong star
formation signposts and sources with extended cm-emission. While the
first group has the steepest indices with a mean around 1.6 the other
two groups have inner radial intensity indices $m_i$ around
1.1. Therefore we propose an evolutionary scenario: star formation
begins with very flat intensity distributions reflecting the initial
conditions in molecular clouds, then during the strongest collapse and
accretion phase the cores are very peaked with steep radial intensity
distributions similar to those known from low mass scenarios, and
finally when the first massive stars have ignited they inject so much
energy into the surrounding molecular material that the cores disrupt
and the intensity distributions flattens out again. Further support of
this scenario is given by the C$^{34}$S linewidth which broadens for
the hot core type sources significantly.

Deriving density distributions from the inner radial indices we get
density power laws around $n \sim r^{-1.6\pm 0.5}$, which is not
very different from values found in low mass-cores \citep{shu
  1977,motte 2001}. But regarding the large spread and the
difficulties to constrain those density parameters tightly, stronger
interpretations of the derived density indices seem to be at least
questionable.

CS and C$^{34}$S observations were undertaken in different rotational
lines. An LVG analysis leads to CS column density around
$10^{15}$~cm$^{-2}$ and local gas densities around a few times
$10^6$~cm$^{-3}$. The latter is higher than the average densities
derived from the mm-continuum emission. This is expected because
inhomogeneous density distributions, clumping and fragmentation of
molecular gas is known to be an ubiquitous phenomenon in molecular
clouds.

Comparing linewidth-mass relations of our sample with more evolved
ultracompact H{\sc ii} regions from the literature indicate that
massive cores are near virial equilibrium before the collapse phase,
then with starting star formation they leave the virial equilibrium
track and collapse violently until the first massive stars ignite
which starts to stabilize the cores again at least partly.

\acknowledgements
We are indebted to Robert Zylka for providing and helping with the
MOPSI software.  We like to thank Frank Bertoldi a lot, who did most
of the MAMBO observations at Pico Veleta. Additionally, we thank Bernd
Weferling for conducting parts of the MAMBO observations as well. Data
of ultracompact H{\sc ii} regions were kindly provided by Jenny
Hatchell. H. Beuther gets support by the {\it Deutsche Forschungsgemeinschaft,
DFG} project number SPP 471. F. Wyrowski is supported by the National
Science Foundation under Grant No. AST-9981289.

\begin{deluxetable}{lrrrr}
\tablecaption{Observing parameters \label{obs}}
\startdata
line & freq  & HPBW   & T$_{sys}$ & resolution \\
     & [GHz] & [$''$] & [K]       & [km/s] \\ 
\tableline
CS 2--1 & 97.98 & 27 & 120 & 0.5 \\ CS 3--2 & 146.97 & 17 & 380 & 0.3
\\ CS 5--4 & 244.94 & 11 & 570 & 0.4 \\ C$^{34}$S 2-1 & 96.41 & 27 &
370 & 0.5 \\ C$^{34}$S 3-2 & 144.62 & 17 & 970 & 0.3 \\ C$^{34}$S 5-4
& 241.02 & 11 & 1220 & 0.2 \\ \enddata\\ \end{deluxetable}

\begin{deluxetable}{lrrrrrrrrr}
\tablecaption{Fits to the 1.2~mm observations \label{mm1}}\tabletypesize{\tiny}
\startdata
source & \# & peak           & int. flux & $\Delta$x & $\Delta$y & maj    & min    & PA & type \\ 
       &    & mJy/$(11'')^2$ & [Jy]      & $['']$    &   $['']$  & $['']$ & $['']$ &    & \\
\tableline
   05358+3543  &  1  &    887 (   19 ) &  6.0  &    27.5 (  0.3 ) &    33.3 (  0.3 ) &   26.7 (  0.6 ) &   19.8 (  0.8 ) &     45   & 1  \\ 
               &  2  &    272 (    5 ) &  0.6  &     5.5 (  0.3 ) &    45.7 (  0.5 ) &   46.0 (  0.8 ) &   18.1 (  2.4 ) &     31   & 0 \\ 
               &  3  &    166 (    2 ) &  0.4  &   $-$19.8 (  0.6 ) &    83.9 (  0.6 ) &   47.2 (  0.5 ) &   19.4 (  2.8 ) &     45 & 0  \\ 
               &  4  &    231 (    6 ) &  1.0  &     8.8 (  0.6 ) &    $-$0.8 (  0.6 ) &   33.6 (  2.0 ) &   29.6 (  1.6 ) &     45 & 0 \\ 
   05490+2658  &  1  &    176 (    3 ) &  1.3  &    $-$7.8 (  0.4 ) &     4.4 (  0.8 ) &   43.9 (  1.0 ) &   27.8 (  2.4 ) &      5 & 3 \\ 
               &  2  &    186 (    5 ) &  1.1  &   $-$17.3 (  0.6 ) &    58.8 (  0.7 ) &   48.3 (  1.7 ) &   34.2 (  2.5 ) &     21 & 3  \\
\enddata\\
\tablecomments{Presented are the peak flux, the integrated flux, the positional offsets, the major and minor axes of the 2-dimensional Gaussian fits and the position angle. The last column shows the three sub-groups as outlined in \S \ref{subgroups}. 0: no grouping, 1: CH$_3$OH and/or CH$_3$CN detections and no resolved cm-source, 2: resolved cm-emission, 3: no CH$_3$OH and CH$_3$CN emission and no resolved cm-source \citep{sridha}}
\end{deluxetable}

\begin{deluxetable}{lrrrrrrrrrrrrrr}
\tablecaption{Results from the 1.2~mm and CS data \label{mm2}}\tabletypesize{\tiny}
\startdata
source & \# & $M_{\rm{far}}$ & $M_{\rm{near}}$ & N$_{\rm{gas}}$      & A$_V$ & $n_{\rm{far}}$    & $n_{\rm{near}}$ & m$_i$ & m$_o$ & b & $p$ & N(CS) & n$_{40K}$ & N$_{40K}$(CS) \\ 
\tableline
   05358+3543  &  1  &    613 &    -- & 5.8 &  617 & 9.3 &  -- &  1.4  &   --  &   5.5  &   7.0$-$7.5  &  14.25$-$15.0  &   5.5$-$6.0  &  14.5$-$14.75  \\ 
               &  2  &     57 &    -- & 1.8 &  191 & 2.1 &  -- &     -- &     -- &     -- &           -- &           -- &           -- &           -- \\ 
               &  3  &     41 &    -- & 1.1 &  117 & 1.2 &  -- &     -- &     -- &     -- &           -- &           -- &           -- &           -- \\ 
               &  4  &    102 &    -- & 1.5 &  159 & 1.8 &  -- &     -- &     -- &     -- &           -- &           -- &           -- &           -- \\ 
   05490+2658  &  1  &    195 &    -- & 1.2 &  127 & 1.1 &  -- &  0.9  &   --  &   6.0  &   7.0$-$7.3  &  13.0$-$14.5  &   5.5$-$6.0  &  13.0$-$14.5  \\ 
               &  2  &    165 &    -- & 1.3 &  138 & 1.0 &  -- &  0.5  &     -- &   3.5  &   6.8$-$7.1  &  13.0$-$14.5  &   5.5$-$6.0  &  13.0$-$14.5  \\ 
   05553+1631  &  1  &    197 &    -- & 1.7 &  180 & 2.4 &  -- &  1.3  &   --  &   0.5  &      $>$7.6  &  13.25$-$15.25  &  6.25$-$6.5  &  13.25$-$14.75  \\
\enddata\\
\tablecomments{The units are: [M$_{\odot}$] for $M_{\rm{far}}$ and $M_{\rm{near}}$, [$10^{23}$cm$^{-2}$] for N$_{\rm{gas}}$, [$10^5$cm$^{-3}$] for  $n_{\rm{far}}$ and  $n_{\rm{near}}$, [log(cm$^{-3}$K)] for $p$, [log(cm$^{-2}$/(km s$^{-1}$))] for N(CS) and N$_{40K}$(CS), and [log(cm$^{-3}$)] for n$_{40K}$.}
\end{deluxetable}

\begin{deluxetable}{lrrrrrrrrrrrrrrrrrrrr}
\tablecaption{Parameters of the CS-fits   \label{table_cs}}\tabletypesize{\tiny}
\startdata
1 & 2 & 3 & 4 & 5 & 6 & 7 & 8 & 9 & 10 & 11 & 12 & 13 & 14 & 15 & 16 & 17 & 18 & 19 & 20 & 21 \\ 
\tableline
 05358+3543  & 1 &   -16.3 &    6.5 &   35.9 &    5.1 &    9.2 &   47.5 &    4.8 &    9.5 &     39
                &    3.8 &     -- &     -- &     -- &     -- &     -- &     -- &     -- &     -- &     -- \\ 
 05490+2658  & 1 &     0.7 &    2.2 &    3.9 &    1.6 &    2.8 &    5.1 &    1.7 &    1.6 &    3.3
                &    1.8 &     -- &     -- &     -- &     -- &     -- &     -- &     -- &     -- &     -- \\ 
           & 2 &     0.3 &    3.3 &    8.9 &    2.5 &    3.4 &    9.8 &    2.7 &    2.3 &    6.1
                &    2.4 &     -- &     -- &     -- &     -- &     -- &     -- &     -- &     -- &     -- \\ 
 05553+1631  & 1 &     5.7 &    2.7 &    6.2 &    2.1 &      4 &   10.6 &    2.4 &    5.1 &     12
                &    2.2 &     -- &     -- &     -- &     -- &     -- &     -- &     -- &     -- &     -- \\ 
 18089-1732  & 1 &    34.1 &    3.7 &     22 &    5.5 &    6.5 &   50.7 &    7.3 &   13.6 &   97.3
                &    6.7 &    3.1 &   12.2 &    3.6 &    6.4 &   27.2 &    3.9 &      8 &     37 &    4.3 \\ 
           & 4 &    34.1 &    2.1 &   10.1 &    4.5 &      3 &     15 &    4.5 &     -- &     --
                &     -- &    0.4 &    1.2 &    2.6 &    0.6 &    1.4 &    2.2 &     -- &     -- &     -- \\ 
 18090-1832  & 1 &   110.8 &    1.4 &    8.9 &    5.9 &    1.6 &   12.9 &    7.4 &    1.3 &    9.1
                &    6.3 &    0.2 &    1.2 &    5.3 &    0.3 &    1.5 &      4 &     -- &     -- &     -- \\
\enddata\\
\tablecomments{The first three columns show the source name, the peak number and the $v_{LSR}$ (as taken from the CS 2--1 fit), the following 18 columns present the peak temperatures  $T_{mb}$, the integrated intensities $\int T_{mb}dv$ and the line widths $\Delta v$ of CS 2--1 (columns 4-6), CS 3--2 (columns 7-9), CS 5--4 (columns 10-12), C$^{34}$S 2--1 (columns 13-15), C$^{34}$S 3--2 (columns 16-18), and C$^{34}$S 5--4 (columns 19-21), respectively.}
\end{deluxetable}

{\bf Figures 1-6:} MAMBO 1.2~mm continuum maps. The axis show offsets in arcsec from the absolute {\it IRAS}-positions given in \citet{sridha}. The greyscale and contours are chosen to highlight the most important features of each image, usually from $10\%$ to $90\%$ (step $10\%$) of the peak flux given in Table \ref{mm1}. \label{figures}

{\bf Figures 7-9:} Example-fits of the
radial profiles: an outer power law, and inner power law and an inner
breakpoint where the power law stops and merges into a constant flux
distribution. \label{fits} 

{\bf Figure 10} {\bf (a)} Histogram of inner radial index $m_i$ 
($I \propto r^{-m_i}$) for all sub-clumps, where fits were
possible. {\bf (b)} $m_i$ histogram for different sub-groups:
grey: strong in hot core tracer CH$_3$OH and/or CH$_3$CN and no
resolved cm-emission; dotted line: resolved cm-sources; dashed line:
CH$_3$OH and CH$_3$CN non-detections and no resolved cm-emission. {\bf
(c)} Histogram of the differences between $m_o$ and
$m_i$. \label{histo_profiles}

{\bf Figure 11:} Results of the
CS-LVG modeling: histograms of gas density distribution {\bf (a)} and
CS column density distribution {\bf (b)}. \label{histo_cs}

{\bf Figure 12:} {\bf(a)} shows
the linewidth correlation between C$^{34}$S(2--1) and NH$_3$. While
filled and open circles represent $\Delta v$(NH$_3(1,1)$) and $\Delta
v$(NH$_3(2,2)$), respectively, the asterisks and open triangles show
$\Delta v$(C$^{34}$(2-1)) versus $\Delta v$(NH$_3(1,1)$) and $\Delta
v$(NH$_3(2,2)$) data for a few UCH{\sc ii} regions \citep{churchwell
1990,cesaroni 1991}. In Figure {\bf (b)} the C$^{34}$S(2--1) linewidth
are plotted versus the core mass (again from the dust
continuum). Filled circles show sources from our sample, while
asterisks and open squares represent data for UCH{\sc ii} regions (for
details see the main text). The full line describes the expected
relation in case of virial equilibrium. In both plots only sources
without distance ambiguity (see
\citet{sridha}) are included. \label{widths}

{\bf Figure 13:} Linear breakpoint of the radial fits versus the
corresponding core masses for sources without distance ambiguity. The
straight line corresponds to $\rm{breakpoint}_{linear} \sim
M^{0.6}$. The sources shown as open circles were not included in the
fit. \label{break_mass}

\end{document}